\documentclass[runningheads]{llncs}
\usepackage{graphicx}
\usepackage{makeidx}  
\usepackage{amssymb}
\usepackage{amsmath}
\usepackage{graphicx}

\usepackage{url}
\usepackage{cite}
\usepackage{type1cm}
\usepackage{eso-pic}
\usepackage{subfigure}
\usepackage[plain,chapter]{algorithm}
\usepackage{algorithmic}

\usepackage{enumerate}
\usepackage[verbose]{placeins}

\usepackage{color}

\begin{document}
\title{Knowledge Discovery on Blockchains:\\Challenges and Opportunities for Distributed Event Detection under Constraints\thanks{Supported by German Research Foundation DFG under grant SFB 876 ''Providing Information by Resource-Constrained Data Analysis`` project B4 ''Analysis and Communication for Dynamic Traffic Prognosis``.}}
\titlerunning{Knowledge Discovery on Blockchains}
%
\author{Cedric Sanders\inst{1}\and
Thomas Liebig\inst{1,2,3}}
\authorrunning{C. Sanders and T. Liebig}
%
\institute{$^1$TU Dortmund, Artificial Intelligence Unit, 44269 Dortmund, Germany
\email{cedric.sanders@tu-dortmund.de},
\email{thomas.liebig@tu-dortmund.de}\\
\url{http://www-ai.cs.uni-dortmund.de/index.html}\\
$^2$Universtiy of Nicosia, Artificial Intelligence Unit, PO 24005 CY-1700 Nicosia, Cyprus\\
$^3$Materna Information \& Communications SE, Artificial Intelligence Unit, 44141 Dortmund, Germany
}
\maketitle              
\begin{abstract}
We study the applicability of blockchain technology for distributed event detection under resource constraints. Therefore we provide a test-suite with several promising consensus methods (Proof-of-Work, Proof-of-Stake, Distributed Proof-of-Work, and Practical Proof-of-Kernel-Work). \\
This is the first work analyzing the communication costs of blockchain consensus methods for knowledge discovery tasks in resource constraint devices. The experiments reveal that our proposed implementations of Distributed Proof-of-Work and Practical Proof-of-Kernel-Work provide a benefit over Proof-of-Work in CPU usage and communication costs.
The tests show further that in cases of low data rates, where latencies by mining do not cause harm proposed blockchain implementations could be integrated. 
However, usage of blockchain requires data broadcasts, which leads to communication overhead as well as memory requirements based on the address list.

\end{abstract}
\section{Introduction}
\label{sec:intro}
The current shift towards edge analysis and distributed knowledge discovery \cite{lazerson18,kamp2018efficient} is mostly driven by making use of large computation clusters and the internet of things. Indeed, applications that benefit from decentralized data management and analysis are, 
amongst others, sensor networks and mobility based services. In both scenarios, a potentially large number of heterogeneous devices is connected and forms a system. The differences amongst the devices could be vast: computation power, memory limitations, energy consumption, etc. Besides, possible applications pose varying requirements for data management. While security is more critical in case of processing vulnerable private information (e.g., medical data), memory consumption or power consumption could be more essential for other application domains.

Once the sensor data in the mesh should be analyzed, one faces the challenge of how to store the data distributedly and how to perform the analysis on this data. This also incorporates the problem of keeping the information amongst the devices consistent. A possible technology that might provide a solution to these issues is the blockchain. These are sequences of unbreakably linked tuples, so-called blocks, of data, transactions, timestamp and the hash value of the ancestral block. A consensus method is required to extend such a blockchain;  this is a procedure how multiple network participants find a new block which is added to the blockchain. Existing consensus methods have requirements in computation costs that ubiquitous devices hardly meet.

Thus, the paper-at-hand fits under the topic of ubiquitous knowledge discovery \cite{may08}. Which connects current advances of data mining and machine learning with the latest developments in internet-of-things and mobile, distributed systems of heterogeneous devices.
This work, therefore, aims at answering the question, whether blockchains are a technical-ready method to process data in distributed heterogeneous networks. We will examine various consensus methods. With well-suited experiments, figures are provided which assist in assessing the general utility of the different blockchain technologies. This raises the following questions:
\begin{itemize}
    \item How should a consensus method operate that meets requirements of cpu usage, memory usage and power consumption originating from a decentralized usage?
    \item Which drawbacks make current consensus method implacable? and How could they be tackled?
    \item Which challenges and requirements remain after analysis of the consensus methods?
\end{itemize}

Many domains for decentralized knowledge discovery could be imagined. Especially citizen science projects, where citizens build sensors and voluntary collect data poses opportunities to distributed immutable knowledge extraction without any centralized coordinator. As a blockchain does not alter the data nor restrict access, the analysis results will not differ from a knowledge discovery in databases \cite{fayyad1996data} or streaming method \cite{domingos2000mining}. However, different consensus methods are more suitable than others. To assess the methods and evaluate communication load, memory consumption, and CPU load, we carry out experiments using publicly available distributed sensor data from opensensemap\footnote{\url{https://opensensemap.org/}}. It is a citizen science project (maintained by the University of M\"unster) and guides the direction for future applications. We use a state-of-the-art event monitoring method, geometric monitoring \cite{sharfman07}, which uses very few communication to monitor a global function. Thus, we propose a fully decentralized application of the previously coordinated geometric monitoring process. As another contribution, we are first to implement and evaluate the initially vague proposal of Distributed Proof-of-Work \cite{cicada}. Thus, we borrow some concepts from the Practical Proof-of-Kernel Work. The latter is included in our software library and (in contrast to previously proprietary implementation) for the first time made available for open public development\footnote{Our sources and link to the data are available at \url{https://bitbucket.org/cedric_sanders/abschlussarbeit/src/master/}.}.


The following second section of the paper presents different works that are related to the presented topic. It is followed by a general introduction of the functionality of a blockchain. The fourth section describes different approaches for achieving consensus in a blockchain and analyzes them by evaluating their advantages and disadvantages. After understanding the different methods it is time to put them to the test in the form of an experiment, which will be evaluated in the fifth section. This evaluation is followed by the last section containing the conclusion of the paper.

\section{Related Work}

While the field of ubiquitous knowledge discovery is established \cite{may08,wolff2009generic} and nowadays receives much attention \cite{lazerson18,kamp2018efficient} not only at major data mining and machine learning symposiums but with the spread of Industry~4.0 and internet-of-things also in application domains, just a few works focus on the chances a decentralized immutable storage of data could have for knowledge discovery and information retrieval. One famous exception is the application with health care data \cite{griggs2018healthcare}, which focuses on automated distributed monitoring of patients. 

Another highlight was the recent initial coin offering of a machine learning blockchain  \cite{decentralizedml}. The authors offered a market space for algorithms and data, based on smart contracts, but it lacked balancing the workload with a smart consensus method. In the following we briefly describe how a blockchain operates.


\section{Blockchain Fundamentals}

In the following, we give a brief introduction to the blockchain technology. Hash functions will play an important role in the next sections. Thus it is important to recall that those are one-way functions which are easy to compute but hard to reverse. A common choice for such a hash function is SHA256 \cite{pub2012secure}. This hash function is a combination of bitwise logical functions (AND, OR, XOR) and shifts (LSHIFTS, RSHIFTS), for the details, we refer the interested reader to the secure hash standard definition in \cite{pub2012secure}. The bitwise manipulation is part of the basic instruction set of most computer chips nowadays, this speeds computation up. Another important property of these hash functions is to map different input most likely to different output\footnote{We are aware that by reduction of dimensionality collisions must occur, but as the hash function is hard to reverse also the collisions are hard to find.}. 

Blockchains first gained attention with the publication of the white paper ''Bitcoin: A Peer-to-Peer Electronic Cash System`` of Satoshi Nakamoto \cite{sn08}. The blockchain is described as a data structure which consists of smaller elements, the so-called blocks. A block comprises of
\begin{enumerate}
    \item data\footnote{Due to the strong connection to cryptocurrencies often named transactions.}: contains the actual observations (e.g., transactions or sensor readings),
    \item timestamp: is used to define a temporal order on blocks, 
    \item hash: hash value of the previous block.
\end{enumerate}

Every block contains the hash of the previous block, which in turn holds the hash of its predecessor. In case one of the old blocks is modified it is simple to recognize in future blocks as the hash value will not fit the one stored previously. 
To use this data structure in a decentralized network, a consensus method has to be added. 

\section{Consensus Methods}

The consensus is an essential part of distributed systems. With blockchains, consensus methods are the class of algorithms that describe how multiple parties find consent on blocks and which novel blocks are added to the chain. Nakamoto describes in his work \cite{sn08} Proof-of-Work, which is still in use nowadays in Bitcoin, as one of these methods. 
In the meantime a bunch of new methods was introduced, for example, Proof-of-Burn \cite{pob}, Proof-of-Luck \cite{pol}, Proof-of-Stake \cite{king12}, and Proof-of-Authority. Ethereum (a distributed platform empowering developers to develop blockchains on existing infrastructure) uses a slightly modified version of Proof-of-Work that may cope with large memory requirements of the participating devices \cite{bv14} \cite{ethereum}.

A recent development with the potential to solve the current problem is the Distributed Proof-of-Work (vaguely proposed in \cite{cicada}) and Practical Proof-of-Kernel-Work of Xain \cite{xain}.

Cicada \cite{cicada} is a group of programmers that suggest a decentralized democratic system based on unique human identifiers. Their vaguely proposed consensus method, however, has some strong points. In the paper at hand, we are the first implementing and analyzing this consensus method.

The origin of Xain \cite{xain} is a group of  British scientist with a background in machine learning. The primary focus is the improvement of blockchains by online adjusting block mining difficulties using reinforcement learning. Their distributed method to grant temporary access to physical doors was successfully implemented in vehicular prototypes.  

\subsection{Proof-of-Work}
\label{sec:pow}
As described beforehand, Proof-of-Work (compare Algorithm~\ref{alg:pow}) is the original concept for consensus on a blockchain, introduced in \cite{sn08}. The basic idea is that a party has to gain the right to publish a novel block. This is done by proving that he spent work in terms of computational power for the generation of the block. The proof is enabled by requiring the answer to a complex problem for publishing a block. Usually, a so-called nonce has to be found, which in combination with the new block has a hash value ending on a specific sequence. By changing the length of this predefined sequence, the hardness of the proof could be adjusted. 

As no party knows in advance which party will add a new block, data needs to be broadcasted to all parties. Parties that aim for publication of a block have to collect the data and combine them in a novel block. Afterward, they could start to find a nonce such that the necessary hash condition is met. Whoever performs these steps fastest may publish the new block. 

It might happen that multiple parties mine the block at the same time, this causes the creation of branches (alternative versions of the blockchain). Proof-of-Work tolerates them for a couple of iterations until one of the branches is longer than its alternatives. As a longer branch represents more computational power, it will be considered correct and alternatives will be deleted. 

Blockchain participants do not need to participate in the mining process. As some incentive, there is a reward in crypto-currencies per block and processing fees for transactions.

\begin{algorithm}
\caption{Mining with Proof-of-Work}
\label{alg:pow}
 \hspace*{\algorithmicindent} \textbf{Input} : Last Block\\
 \hspace*{\algorithmicindent} \textbf{Output} : New Block 
\begin{algorithmic}[1]
\STATE{$\textit{nonce} \gets 0$}
\WHILE{\textit{proof} is not valid}
	\STATE{$\textit{proof} \gets$ \textit{ProofofWork(Last Block, nonce)}}
	\STATE{$\textit{nonce} \gets \textit{nonce}+1$}
\ENDWHILE
\STATE{\textit{RewardMiner()}}
\STATE{\textit{CreateBlock(Last Block, proof)}}
\end{algorithmic}
\end{algorithm}

Proof-of-Work is the oldest consensus method incorporated in this study. And since it is the most popular, it is used most often in literature. Thus, the following techniques were introduced to cope with its challenges. 
Proof-of-Work has been introduced to implement a decentralized cryptocurrency. Therefore, safety, decentralism, and resilience were the most critical aspects of the design. 

In his paper, Nakamoto describes attack scenarios and clarifies that one party needs to be faster in adding blocks to the chain than all others to cause damage. At the same time, he estimates the probability an attacker could do this \cite[11. Calculations]{sn08}. However, this safety has its cost.
The blockchain is safe as long as a majority of the computation power is used for mining novel blocks. This has several properties:
The energy consumption of parties that aim for maintaining consistency is exceptionally high. From the very beginning till a possible shut down. This is a problem especially for potentially small or mobile devices which have limited energy budget.
Another drawback is the distribution of computational power which is directly coupled with the integrity of the chain. The problem is often also called 51\% problem, as any cooperative group of parties holding 51\% of the computational power gains a higher impact on the system and attacks on the integrity get easier. 
In the blockchain, parties may join or leave the network at any time, without any inconsistencies (e.g., duplicate or missing data). Therefore data storage has some redundancy. So blockchains are excellent for applications where memory consumption does not matter, and a high resilience needs to be guaranteed.

The last point is the high communication cost with blockchains in general. The distributed storage blockchain requires lots of communication as broadcasts are necessary for transmission of transactions or data, for distribution of novel blocks, for finding the longest blockchain and resolving branches.

\FloatBarrier
\subsection{Proof-of-Stake}
Proof-of-Stake (compare algorithm~\ref{alg:pos}) is a concept introduced by Sunny King and Scott Nadal, for the PPCoin \cite{king12}. It follows an entirely different approach which is stronger coupled with the application as currency. A party does not need to prove an amount of workload but has to prove it owns a certain amount of the currency. 
For this reason, Proof-of-Stake uses coinage. Coinage describes the ownership of coins over a certain period. If a party owns 100 coins over 10 time slices, he holds coinage 1.000. The coinage sinks if coins are spent.
If a party wants to add a block, it transfers itself coins to reduce its coinage. By this transaction, it gains as a reward a simplification of the mining problem.  
\begin{algorithm}
\caption{Mining with Proof-of-Stake}
\label{alg:pos}
 \hspace*{\algorithmicindent} \textbf{Input} : Last Block\\
 \hspace*{\algorithmicindent} \textbf{Output} : New Block 
\begin{algorithmic}[1]
\STATE{$\textit{coinage} \gets$ \textit{CalculateCoinAge()}}
\STATE{$\textit{investment} \gets$ \textit{Random(coinage)}}
\STATE{\textit{MakeInvestment()}}
\STATE{$\textit{nonce} \gets 0$}
\WHILE{\textit{proof} is not valid}
	\STATE{$\textit{proof} \gets$ \textit{ProofofWork(Last Block, nonce, investment)}}
	\STATE{$\textit{nonce} \gets \textit{nonce}+1$}
\ENDWHILE
\STATE{\textit{RewardMiner()}}
\STATE{\textit{CreateBlock(Last Block, proof)}}
\end{algorithmic}
\end{algorithm}

Proof-of-Stake requires any form of ownership to invest it for gaining impact on the blockchain. This reduces its applicability in practice. For sure application domains could be extended with such a concept (as done with Ether tokens in Ethereum). But if the consensus method does not provide a considerable benefit over the other methods this workaround should be scrutinized.

Implementing this method raises new questions. For example, an investment strategy of tokens for miners to spend their tokens. This provides a vast potential for ubiquitous devices with limited computational power. They could take high investments at a low frequency to optimally use their limited computation capabilities.

\FloatBarrier
\subsection{Distributed Proof-of-Work}
The novel consensus method of cicada is called distributed-Proof-of-Work and bases on Proof-of-Work \cite{cicada}. It structures the mining process into small contests called mining races. In contrast to Proof-of-Work, access to mining is limited. For each mining race, a set of participants is randomly selected. Using, Distributed Proof-of-Work any party needs to take part in mining, making them eligible for mining races.

\begin{algorithm}
\caption{Mining with Distributed Proof-of-Work}
\label{alg:dpow}
 \hspace*{\algorithmicindent} \textbf{Input} : Last Block\\
 \hspace*{\algorithmicindent} \textbf{Output} : New Block 
\begin{algorithmic}[1]
\STATE{$\textit{selected} \gets$ \textit{False}}
\IF{enough nodes selected} 
	\IF{selected}
		\STATE{$\textit{nonce} \gets 0$}
		\WHILE{\textit{proof} is not valid}
			\STATE{$\textit{proof} \gets$ \textit{ProofofWork(Last Block, nonce)}}
			\STATE{$\textit{nonce} \gets \textit{nonce}+1$}
		\ENDWHILE
		\STATE{\textit{RewardMiner()}}
		\STATE{\textit{CreateBlock(Last Block, proof)}}
	\ELSE
		\STATE{Wait for next mining race}
	\ENDIF
\ELSE
	\STATE{$\textit{selected} \gets$ \textit{VerifiableRandomFunction(seed, difficulty)}}
	\IF{selected}
		\STATE{Let the other nodes verify the Verifiable Random Function}
	\ELSE
		\STATE{Wait for other nodes to do the lottery}
		\IF{not enough nodes selected} 
			\STATE{Reduce the difficulty of being selected}
		\ENDIF
	\ENDIF
\ENDIF
\end{algorithmic}
\end{algorithm}

Distributed Proof-of-Work restricts access to the original Proof-of-Work to overcome some of its problems. However, the description of the authors is somewhat vague; for example the selection of the miners which causes difficulties for implementation. In the following section on Practical Proof-of-Kernel Work verifiable random functions (VRF) are applied for that purpose. Thus we borrow the concept also for Distributed Proof-of-Work. VRF is a concept to execute a random function, i.e., a function with an unknown result while the processor is capable of proving to other parties that the obtained value is correct. Variable random functions could be constructed with different cryptographic methods. One example is the approach by  Goldberg \cite{goldberg16} using elliptic curve cryptography. The basic procedure follows these steps \cite{micali99}:
\begin{itemize}
    \item A so-called generator provides public $pk$ and private keys $sk$ to each party.
    \item Given its private key $sk$ and a publicly known seed $x$ a party is capable of computing a random function $f$ which calculates a proof $p$.
    \item Every other party could now verify that the proof is the result of the random function given $p$, $pk$ and $x$.
\end{itemize}

\FloatBarrier
\subsection{Practical Proof-of-Kernel-Work}
Practical Proof-of-Kernel-Work (compare algorithm~\ref{alg:ppokw}) also bases on Proof-of-Work but includes access control. The used methods are more complex than those of Distributed Proof-of-Work.
Three mechanisms control participation in Proof-of-Work:
\begin{itemize}
    \item A whitelist that lists trustable parties.
    \item A set of dynamic rules. For example the creator of the last block could be banned for the next three iterations.
    \item A random selection of parties from the whitelist.  
\end{itemize}
This latter selection routine uses a continuous seed that is embedded in the blockchain. This seed may be used by the parties to perform a lottery based on variable random functions. The selection process not only guarantees a random selection but also prevents others from obtaining any knowledge on the selected parties. This prevents an attacker from performing targetted attacks on parties. The chosen parties, in turn, can prove that they have been selected.
\begin{algorithm}
\caption{Mining with Practical Proof-of-Kernel-Work}
\label{alg:ppokw}
 \hspace*{\algorithmicindent} \textbf{Input} : Last Block\\
 \hspace*{\algorithmicindent} \textbf{Output} : New Block 
\begin{algorithmic}[1]
\STATE{$\textit{selected} \gets$ \textit{False}}
\IF{enough nodes selected} 
	\IF{selected}
		\STATE{$\textit{nonce} \gets 0$}
		\WHILE{\textit{proof} is not valid}
			\STATE{$\textit{proof} \gets$ \textit{ProofofWork(Last Block, nonce)}}
			\STATE{$\textit{nonce} \gets \textit{nonce}+1$}
		\ENDWHILE
		\STATE{\textit{RewardMiner()}}
		\STATE{\textit{CreateBlock(Last Block, proof)}}
	\ELSE
		\STATE{Wait for next mining race}
	\ENDIF
\ELSE
	\IF{Node on Whitelist}
		\IF{\textit{CheckRuleset()}}
			\IF{\textit{VerifiableRandomFunction(seed, difficulty)}}
				\STATE{$\textit{selected} \gets$ \textit{True}}
			\ENDIF
		\ENDIF
	\ENDIF
	\IF{selected}
		\STATE{Let the other nodes verify the Selection}
	\ELSE
		\STATE{Wait for other nodes to do the lottery}
		\IF{not enough nodes selected} 
			\STATE{Reduce the difficulty of being selected}
		\ENDIF
	\ENDIF
\ENDIF
\end{algorithmic}
\end{algorithm}

The access control reduces energy consumption and weakens the 51\% problem described above (compare Section~\ref{sec:pow}) as neither computation nor ownership has an impact on future selection.
Practical Proof-of-Kernel-Work reduces the likelihood of branches as fewer parties participates in mining. 
Storing a whitelist on the blockchain holds potential problems for its scalability. In large networks, this would cause huge memory consumption which restricted memory devices may have issues with. Also, the computation of a verifiable random function possesses challenges to computationally weak devices. 

\section{Experiments}

Especially for the two novel and promising consensus methods Distributed Proof-of-Work and Practical Proof-of-Kernel-Work (compare Sections~4.3 and 4.4) no implementation was available and description was rather vague. Thus, we contribute implementations of these consensus methods. 
With the focus on potentially heterogeneous devices and latest developments of micropython and circuitpython as rapid prototyping 'operating systems` for ultra-low power devices, we picked python as a programming language\footnote{The resulting sources are made publicly available at \url{https://bitbucket.org/cedric_sanders/blockchain-experiments/src/master/}.}. 
To obtain comparable results also the two established consensus methods Proof-of-Work and Proof-of-Stake are part of our library. The use of a blockchain could expect no improvement or drawback, thus we will not report on the performance of geometric monitoring, but on our measures of interest CPU load, memory usage, through-put and communication costs.

In the following, we briefly describe the problems we faced.
The analysis presumes random access to the data. Thus the choice of the algorithm is not crucial for a comparison. We implemented nodes that are processed on a system and hold essential functions:
\begin{itemize}
    \item running transactions,
    \item mining blocks,
    \item syncing the blockchain,
    \item reading a stream of observations from a file and inserting observations to the blockchain,
    \item logging of metrics for the analysis we present next,
    \item running a local model for data analysis.
\end{itemize}

The overall goal is to keep different methods comparable. Thus we did not focus on implementing individual cases but a plain structure of the methods, as described in Section~4. 

All our implementations consist of three building blocks 1) a RESTful server that provides an API to other parties, 2) a part for mining and maintenance of the blockchain and 3) a part that performs the actual calculations. For the basic Proof-of-Work, the difficulty was set to 6 leading zeros, which corresponds to 40 seconds block time and fits quite well to the sampling interval of the opensensemap data\footnote{\url{https://opensensemap.org/}} we use.
The practical Proof-of-Kernel-Work requires the inclusion of the verifiable random functions. We applied the approach of  \cite[Definition 4.1]{goldberg16} for the lottery. Syncing the needed seed amongst the parties, however, caused some unexpected problems as the operations are not atomic and there might already be consensus for a new seed once a node finished mining, we overcame this chance of asynchronicity by relaxing the verification. Thus, we allow also ancestor and of current seed as valid. Another important decision is when the results of the lottery are broadcasted. If the set of miners is published directly (before the actual creation of the block) the other parties could send their data or transactions directly to the selected ones. An alternative method would be to reveal the decision of the lottery after mining of the new block, this requires a broadcast of all observations. The latter techniques would be more secure. It prevents targetted attacks on single nodes. But, as we want to see the potential benefit of the consensus method, We decided for the first option which reduces energy consumption and communication cost. 
Besides, we added a white list that keeps a record of the trustable devices. As soon as a malicious party sends fraudulent blocks to the network, it will be removed from the whitelist. Distributed Proof-of-Work operates similar as Practical Proof-of-Kernel-Work, but the time and memory consuming additions are removed. This includes dynamic rules and whitelist.

To compare the consensus methods on their feasibility. We need to test it with a distributed data analysis task. Usage of the blockchain does not alter the data, neither do the considered consensus methods. A distributed analysis, therefore, produces the same result as without using a blockchain. The application we are aiming for is a distributed monitoring task with multiple sensors. We perform analysis with the well suited geometric monitoring approach \cite{sharfman07} that reduces communication costs and bases on a simple concept. Recent improvements were published in \cite{keren14,lazerson18}. The primary task is that a global threshold function should be monitored without communicating every single data item. The communication is reduced by introducing local threshold conditions which need to be raised to start communication with the coordinator. The coordinator checks the global function and updates the threshold parameters of the parties. Challenge for the application of geometric monitoring is the design of the local conditions. The requirements to local conditions are:
\begin{itemize}
    \item Correctness: As long as all local conditions are below the threshold, also the global threshold is not reached.
    \item Communication efficiency: The number of necessary communications is minimal.
    \item Efficient computation: the calculation required to test local conditions is low.
\end{itemize}
As recently shown in \cite{lazerson18}, finding these local threshold functions is more straightforward when the global function is convex. Then it is sufficient to find a close upper bound for the global function.

In combination with blockchain, the geometric monitoring approach could be applied coordinator free, fully decentralized. Every node has the required information to test global functions.

As described above (compare Section~\ref{sec:intro}), we perform the tests using data from opensensemap\footnote{\url{https://opensensemap.org/}}. This is a network of citizen sensor data consisting of 3909 so-called senseboxes. In general, they are situated around the globe, but mostly they are in Germany. The attributes of each sensor records differ much. While just a few sensors record special features as gamma radiation, temperature and wind speeds are prevalent attributes. To perform tests with geometric monitoring, in this study, we decided for the temperature feature\footnote{The data we apply the method to is obtained in the interval from March, 23rd 2019 till March, 24th 2019, in the WGS 84 box [5.98865807458, 47.3024876979, 15.0169958839, 54.983104153].}. Setting a global threshold on the average temperature is easy.

The experiments were conducted on a cluster of multi-core computers each running a process of a node. We tested for 5, 10, 20 and 40 participants. To validate even larger networks future implementation could make use of MPI or other interprocess communication protocols. Direct test in a distributed sensor mesh is another option but in a fully distributed coordinator free setup analysis of the experiment also requires centralization and eventually clock synchronization. 

We analyze four aspects: communication, mining, memory and CPU usage.

The communication analysis is split into the types, data request, transaction received, coordinating blocks and transactions.
\begin{figure}[h]
	\centering
    \begin{minipage}[t]{0.41\linewidth}
        \centering
        \includegraphics[width=\linewidth]{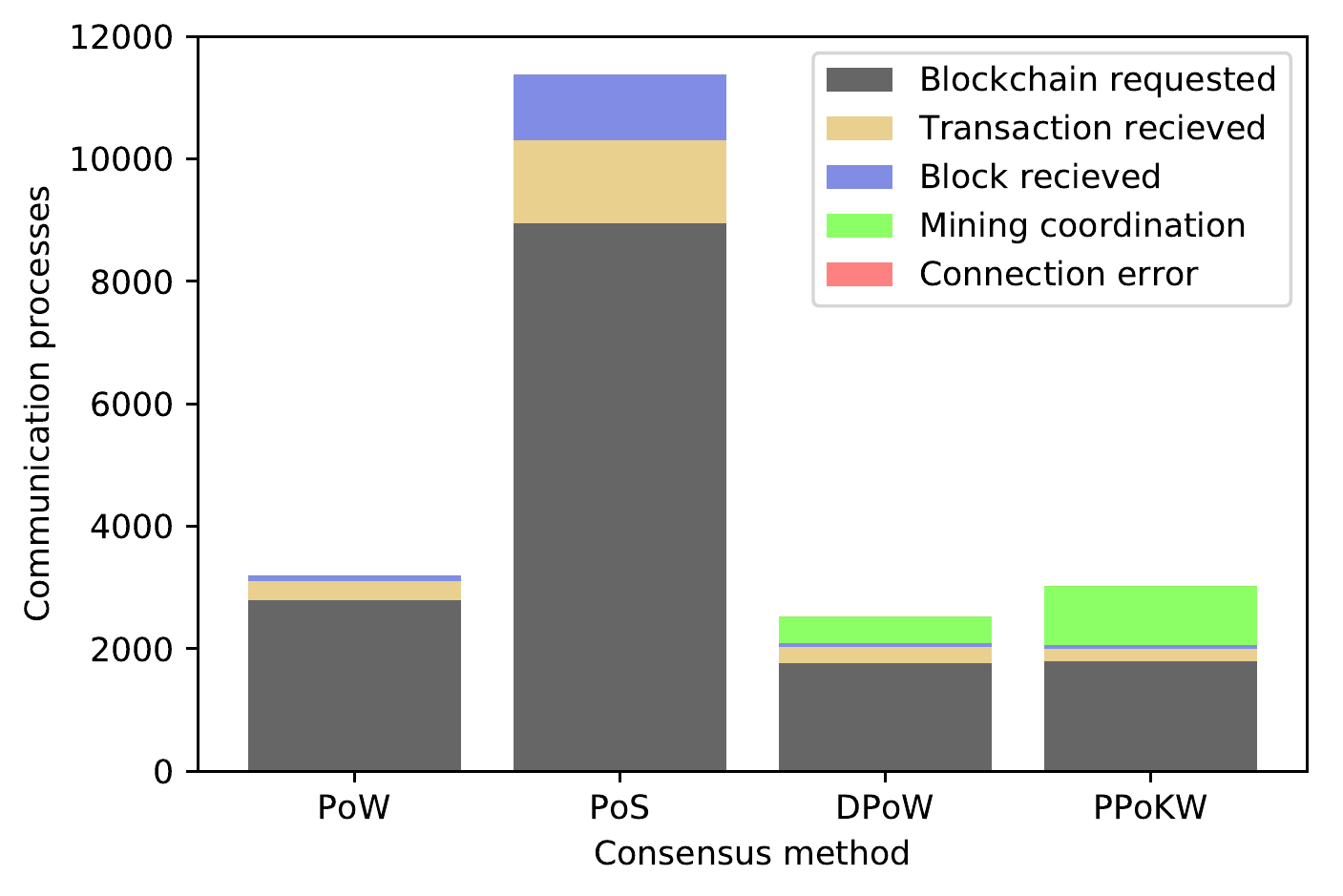}
        \caption{Average communicationcost for 5 nodes over 1 hour}
        \label{fig:comm_5}
    \end{minipage}
    \hfill
    \begin{minipage}[t]{0.41\linewidth}
        \centering
        \includegraphics[width=\linewidth]{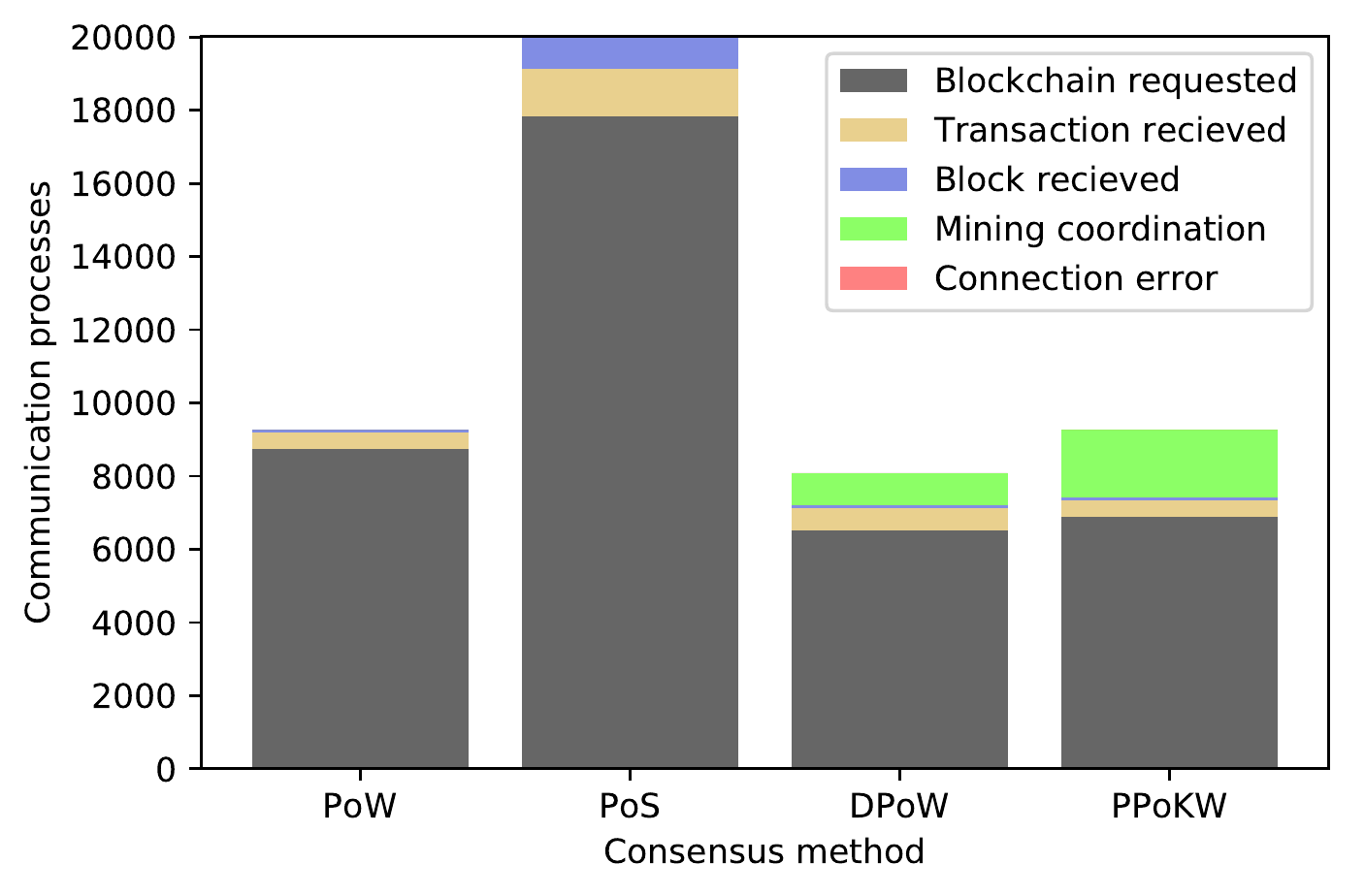}
        \caption{Average communicationcost for 10 nodes over 1 hour}
                \label{fig:comm_10}
    \end{minipage}
	\centering
    \begin{minipage}[t]{0.41\linewidth}
        \centering
        \includegraphics[width=\linewidth]{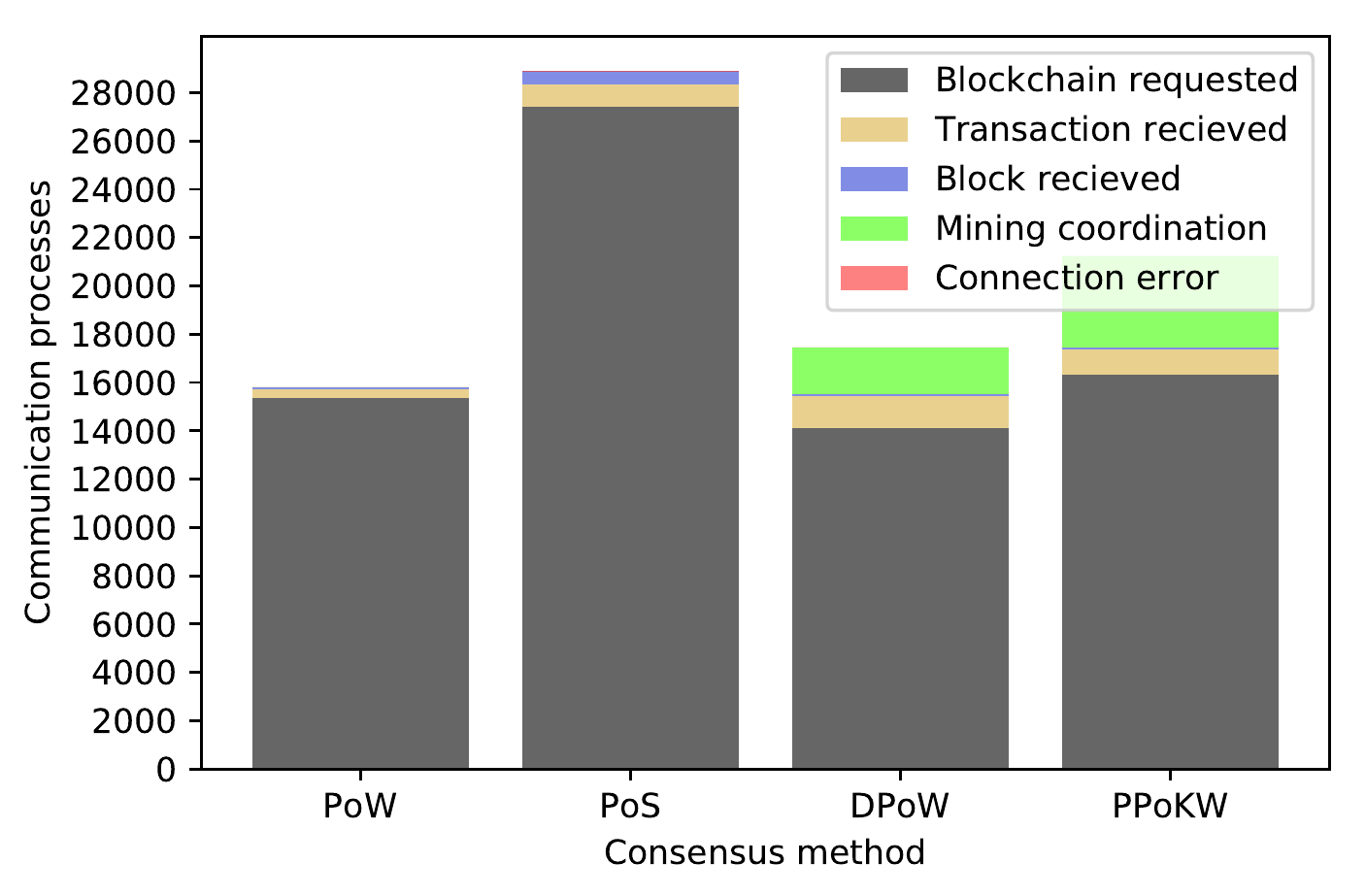}
        \caption{Average communicationcost for 20 nodes over 1 hour}
                \label{fig:comm_20}
    \end{minipage}
    \hfill
    \begin{minipage}[t]{0.41\linewidth}
        \centering
        \includegraphics[width=\linewidth]{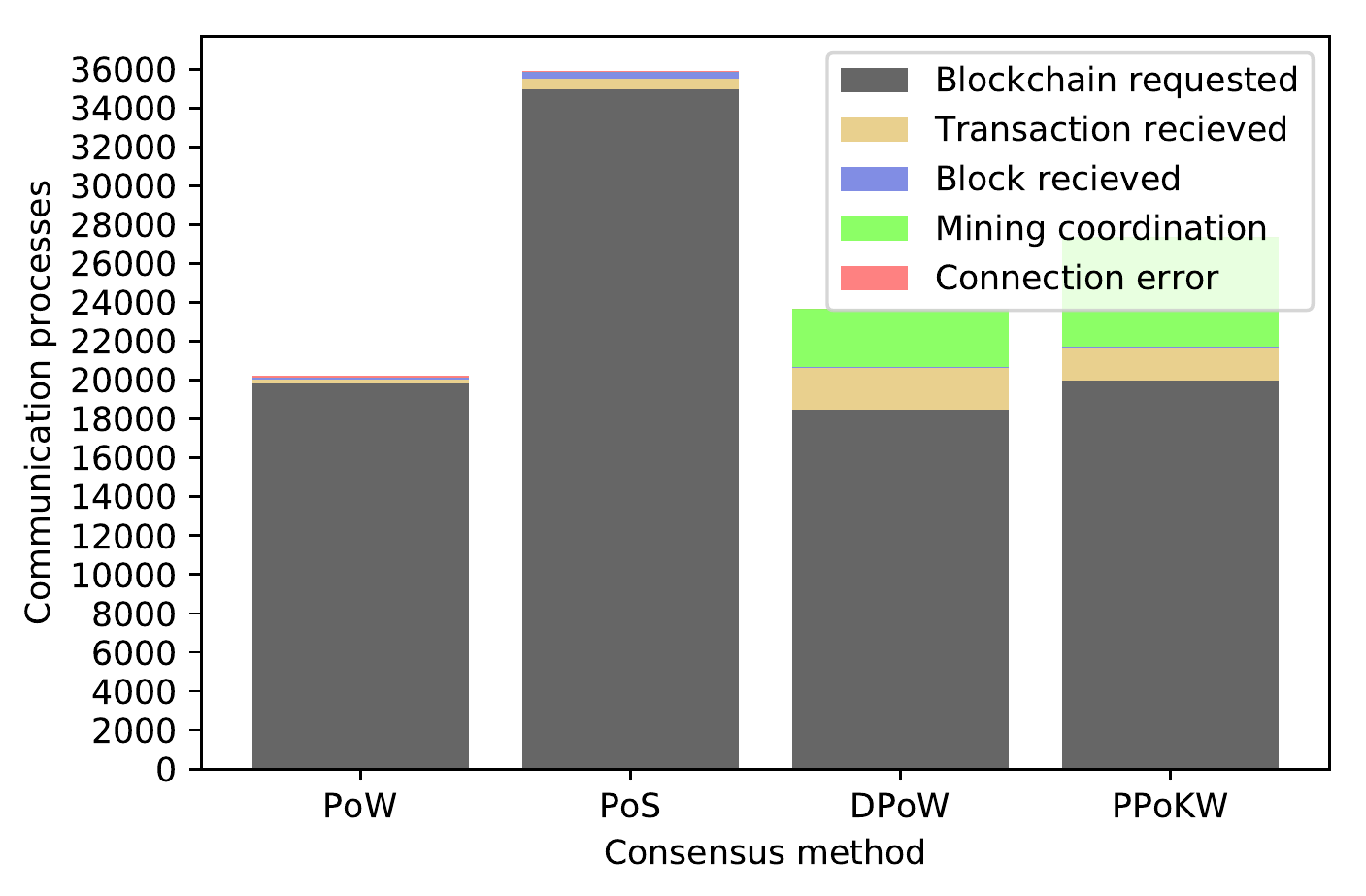}
        \caption{Average communicationcost for 40 nodes over 1 hour}
                \label{fig:comm_40}
    \end{minipage}        
\end{figure}

The Figures~\ref{fig:comm_5} to~\ref{fig:comm_40} reveal that Proof-of-Stake requires more communication than the other methods. One reason is additional transactions to communicate the coinage. The additional checks of the coinage cause more communication rounds on the blockchain. Most communication originates from the access on the blockchain data. The transmission of transactions and blocks are neglectable.

\begin{figure}[h]
	\centering
    \begin{minipage}[t]{0.41\linewidth}
        \centering
        \includegraphics[width=\linewidth]{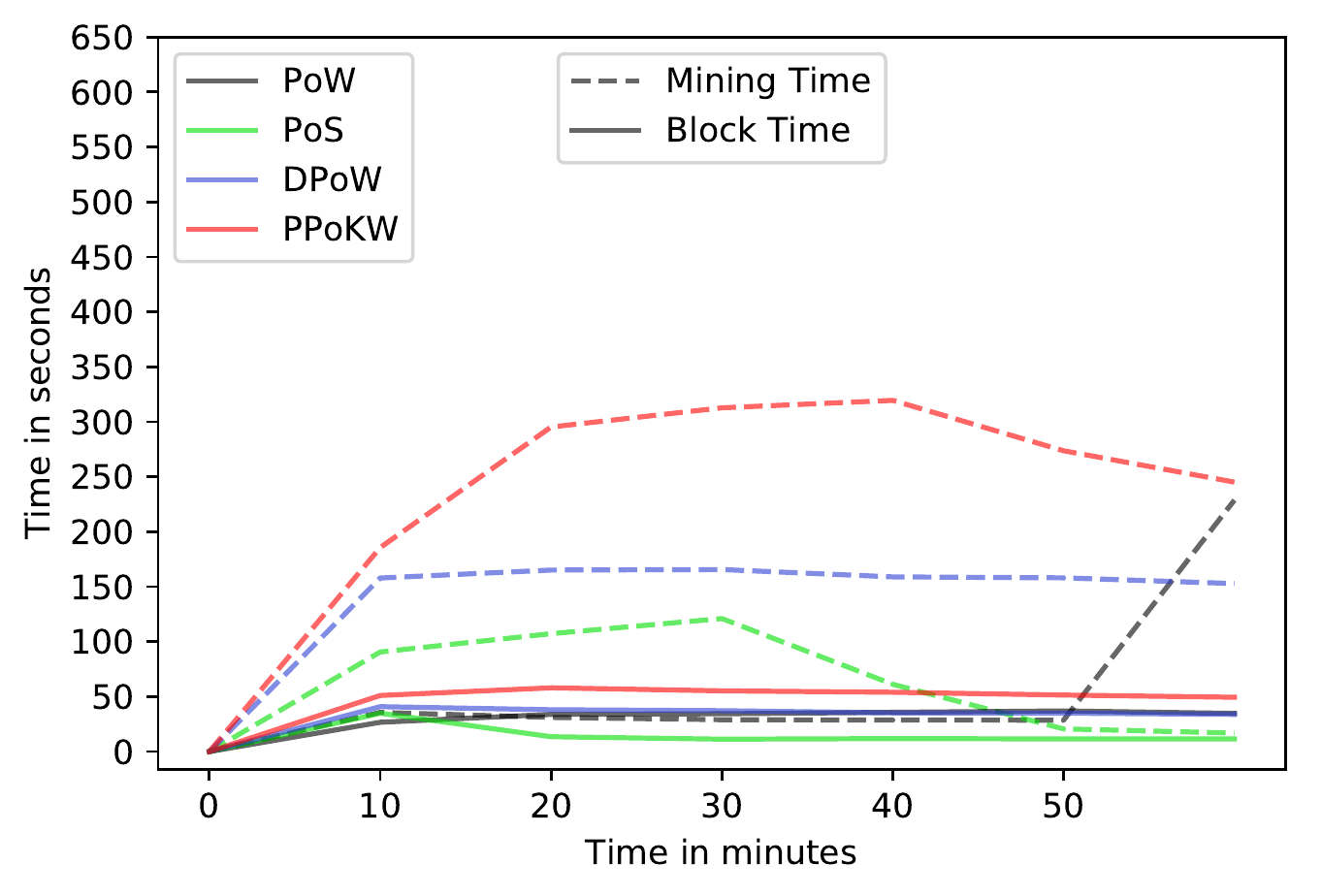}
        \caption{Average Mining Time/ Block Time for 5 nodes over 1 hour}
        \label{fig:blocktime_5}
    \end{minipage}
    \hfill
    \begin{minipage}[t]{0.41\linewidth}
        \centering
        \includegraphics[width=\linewidth]{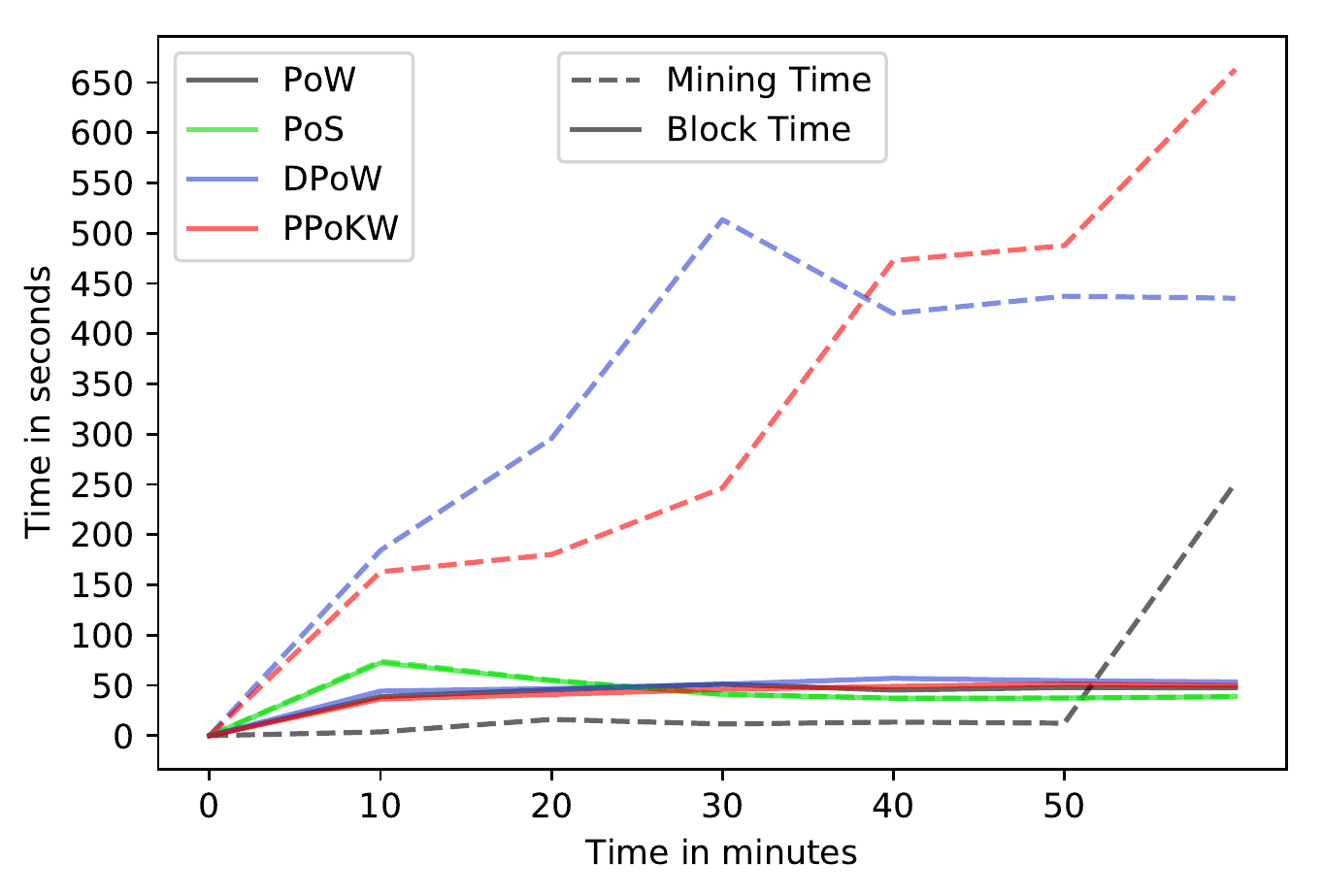}
        \caption{Average Mining Time/ Block Time for 10 nodes over 1 hour}
        \label{fig:blocktime_10}
    \end{minipage}
	\centering
    \begin{minipage}[t]{0.41\linewidth}
        \centering
        \includegraphics[width=\linewidth]{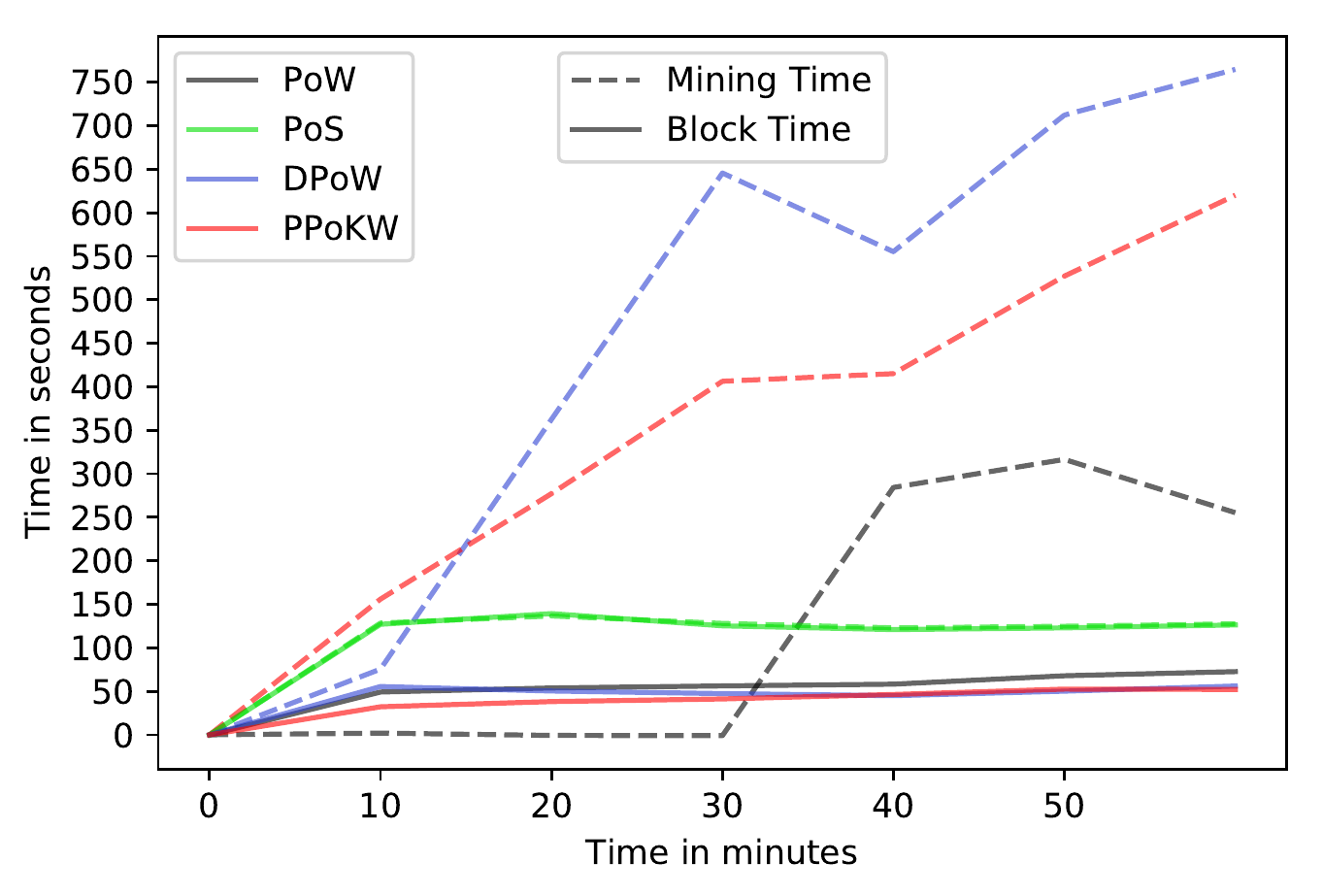}
        \caption{Average Mining Time/ Block Time for 20 nodes over 1 hour}
        \label{fig:blocktime_20}
    \end{minipage}
    \hfill
    \begin{minipage}[t]{0.41\linewidth}
        \centering
        \includegraphics[width=\linewidth]{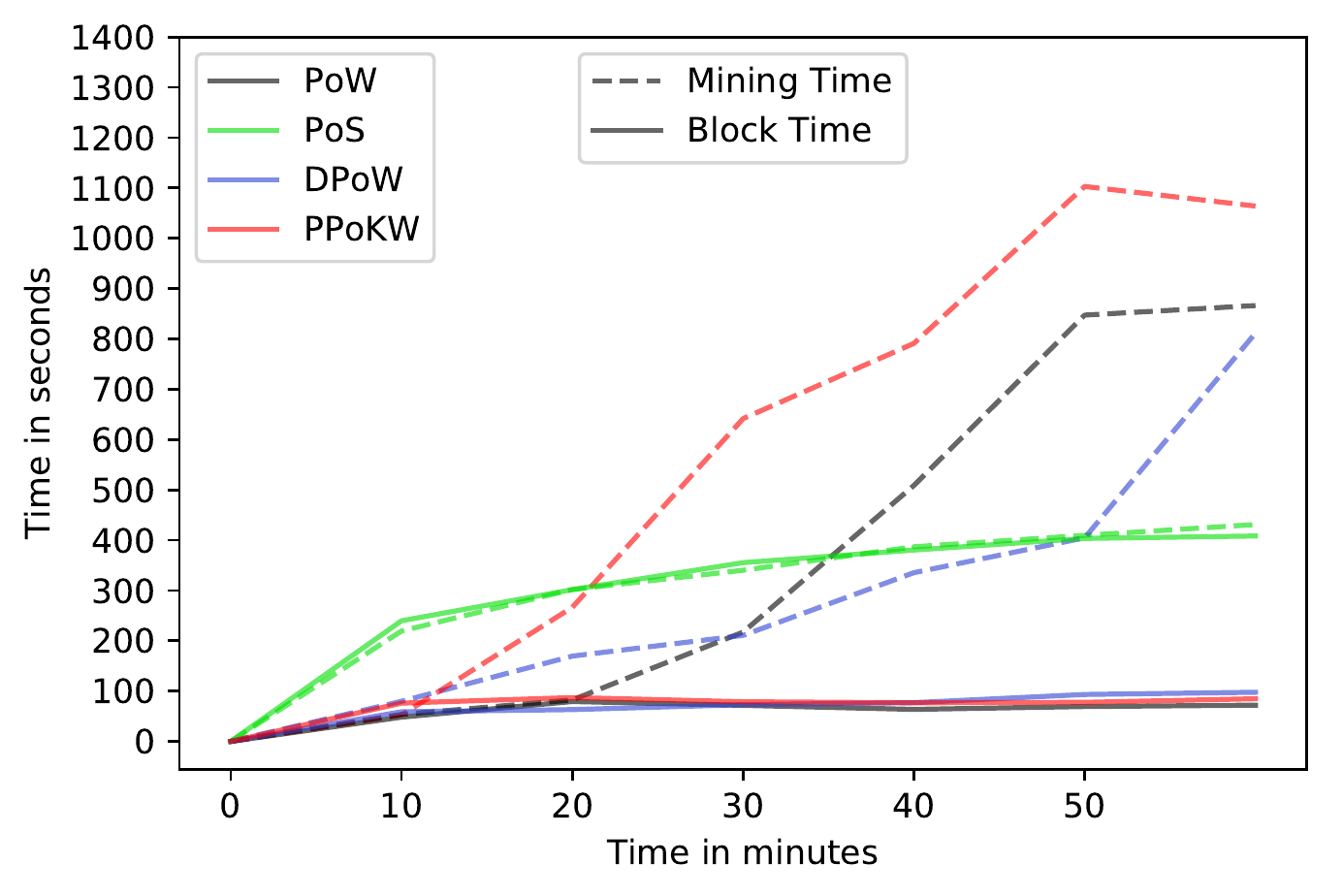}
        \caption{Average Mining Time/ Block Time for 40 nodes over 1 hour}
        \label{fig:blocktime_40}
    \end{minipage}                 
\end{figure}

Next, we test the temporal performance of the blockchain. How much time is required for block generation and whats the block through-put?
\begin{figure}[h]
	\centering
    \begin{minipage}[t]{0.41\linewidth}
        \centering
        \includegraphics[width=\linewidth]{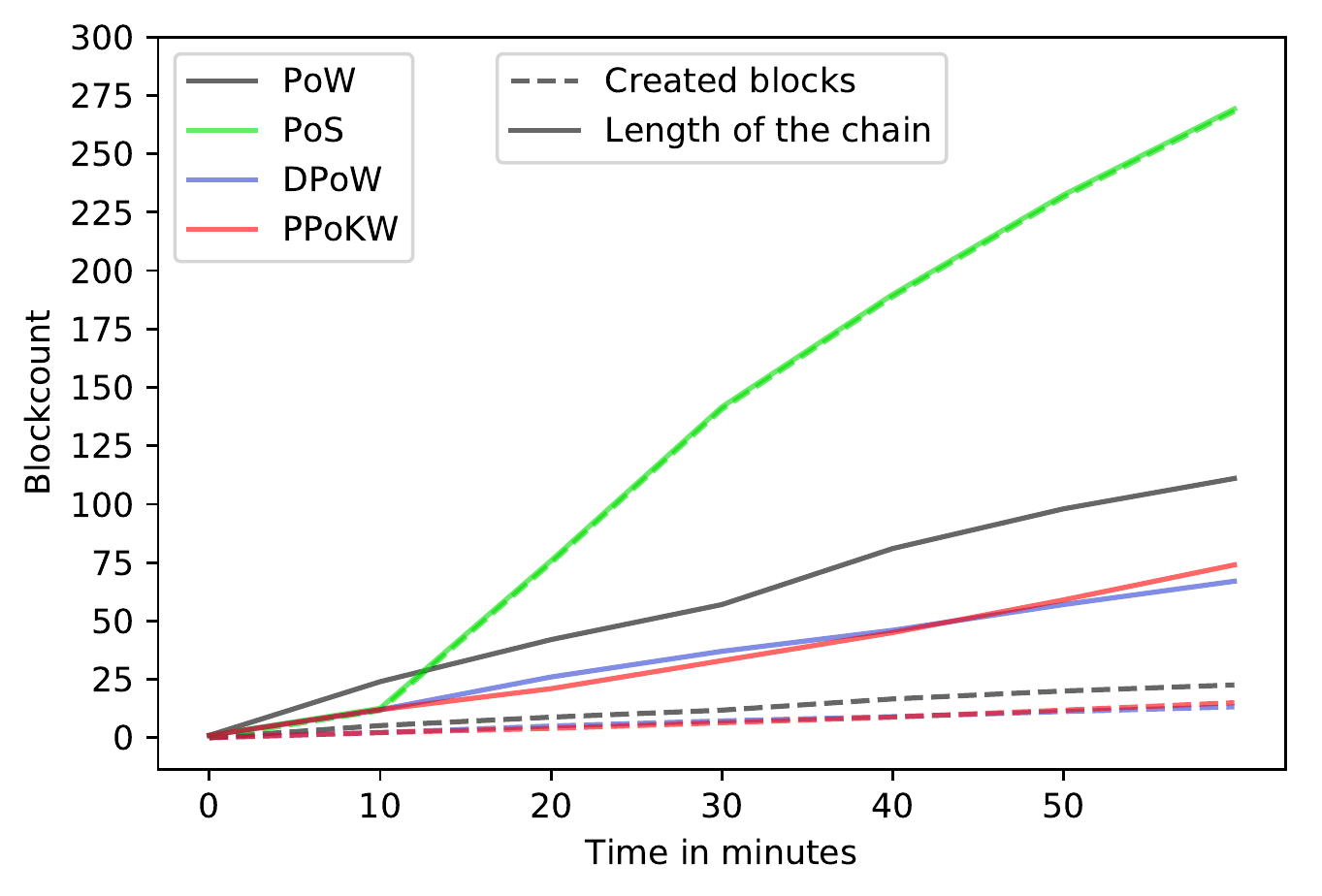}
        \caption{Average created/total blocks for 5 nodes over 1 hour}
        \label{fig:throughput_5}
    \end{minipage}
    \hfill
    \begin{minipage}[t]{0.41\linewidth}
        \centering
        \includegraphics[width=\linewidth]{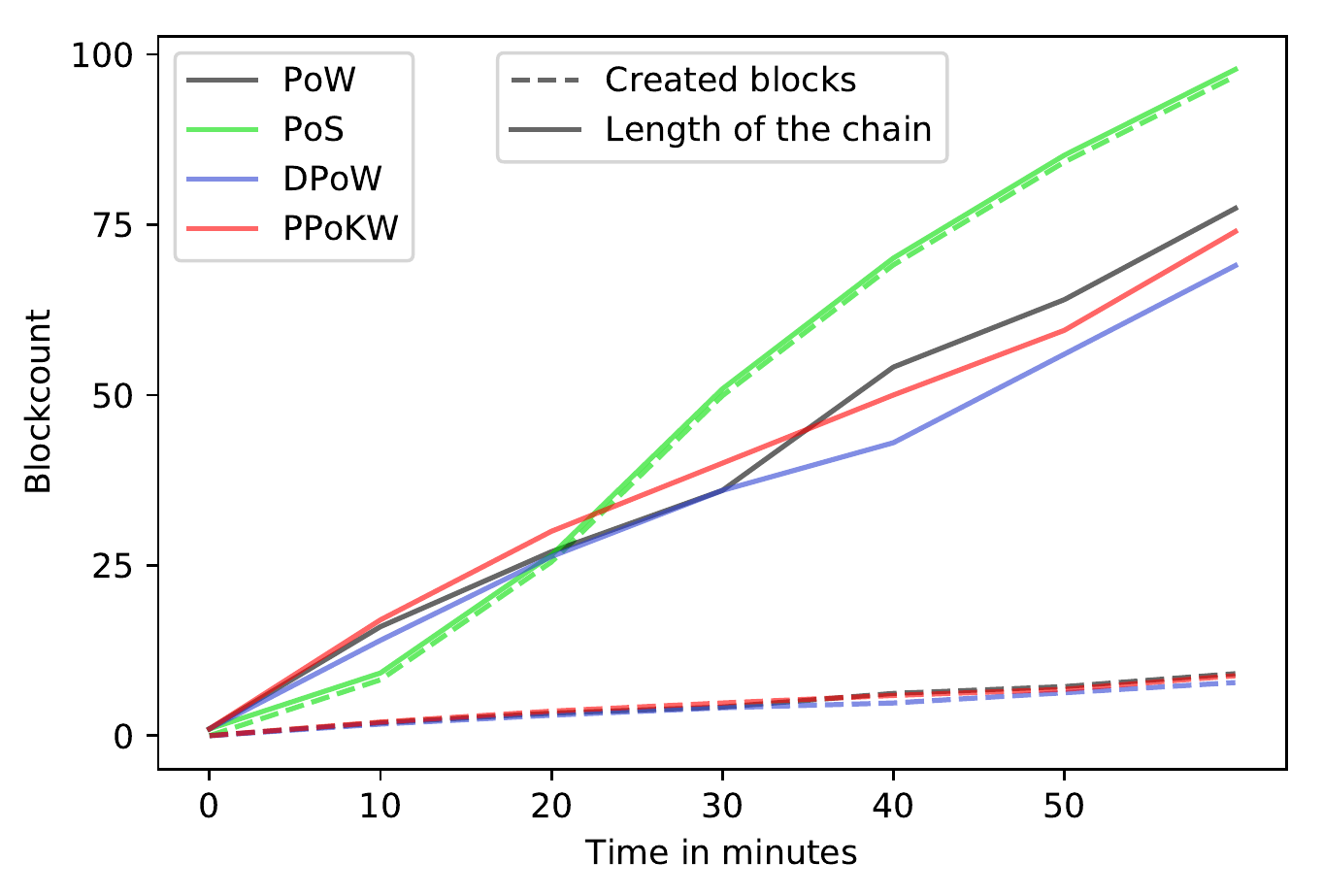}
        \caption{Average created/total blocks for 10 nodes over 1 hour}
        \label{fig:throughput_10}
    \end{minipage}
	\centering
    \begin{minipage}[t]{0.41\linewidth}
        \centering
        \includegraphics[width=\linewidth]{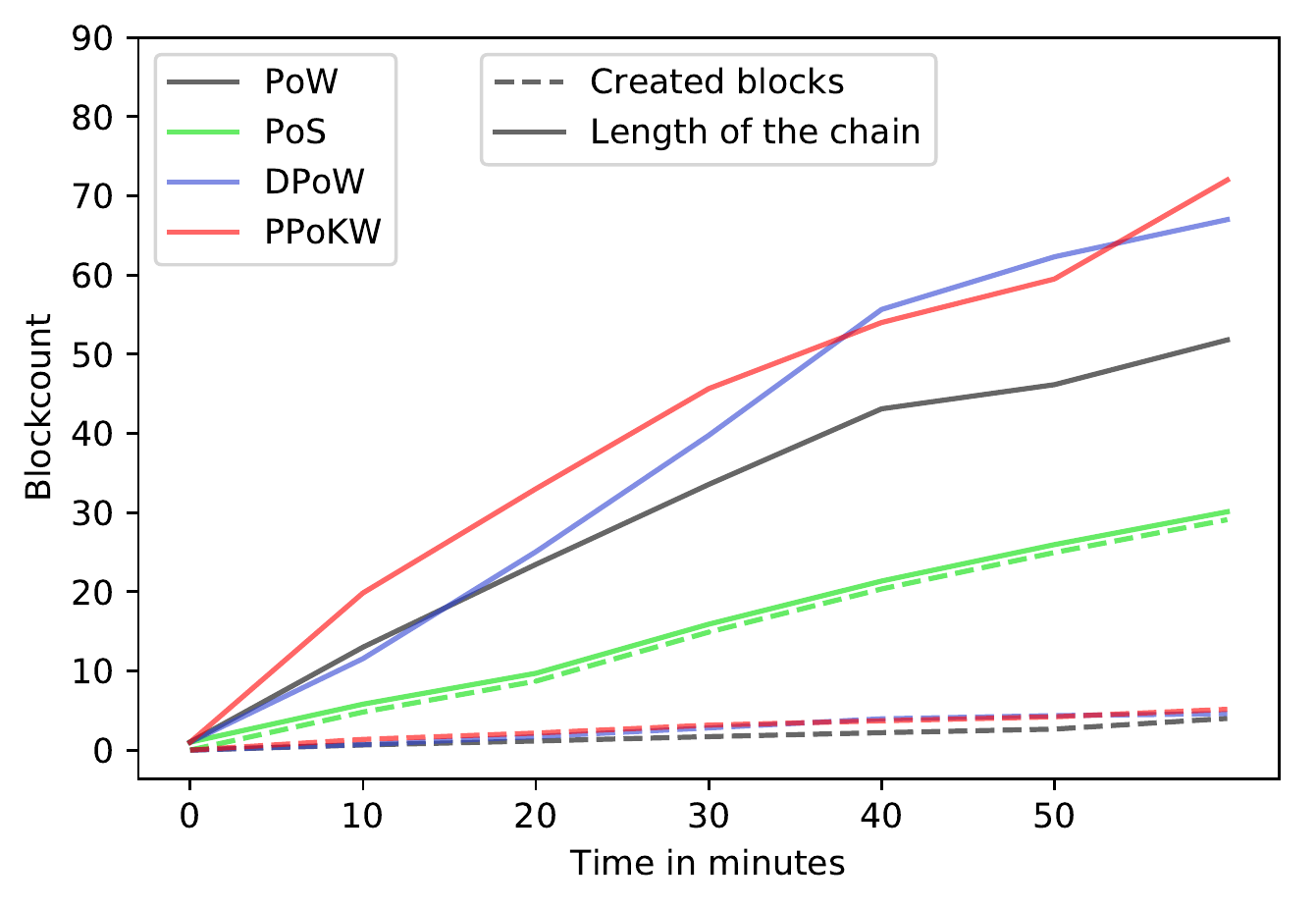}
        \caption{Average created/total blocks for 20 nodes over 1 hour}
        \label{fig:throughput_20}
    \end{minipage}
    \hfill
    \begin{minipage}[t]{0.41\linewidth}
        \centering
        \includegraphics[width=\linewidth]{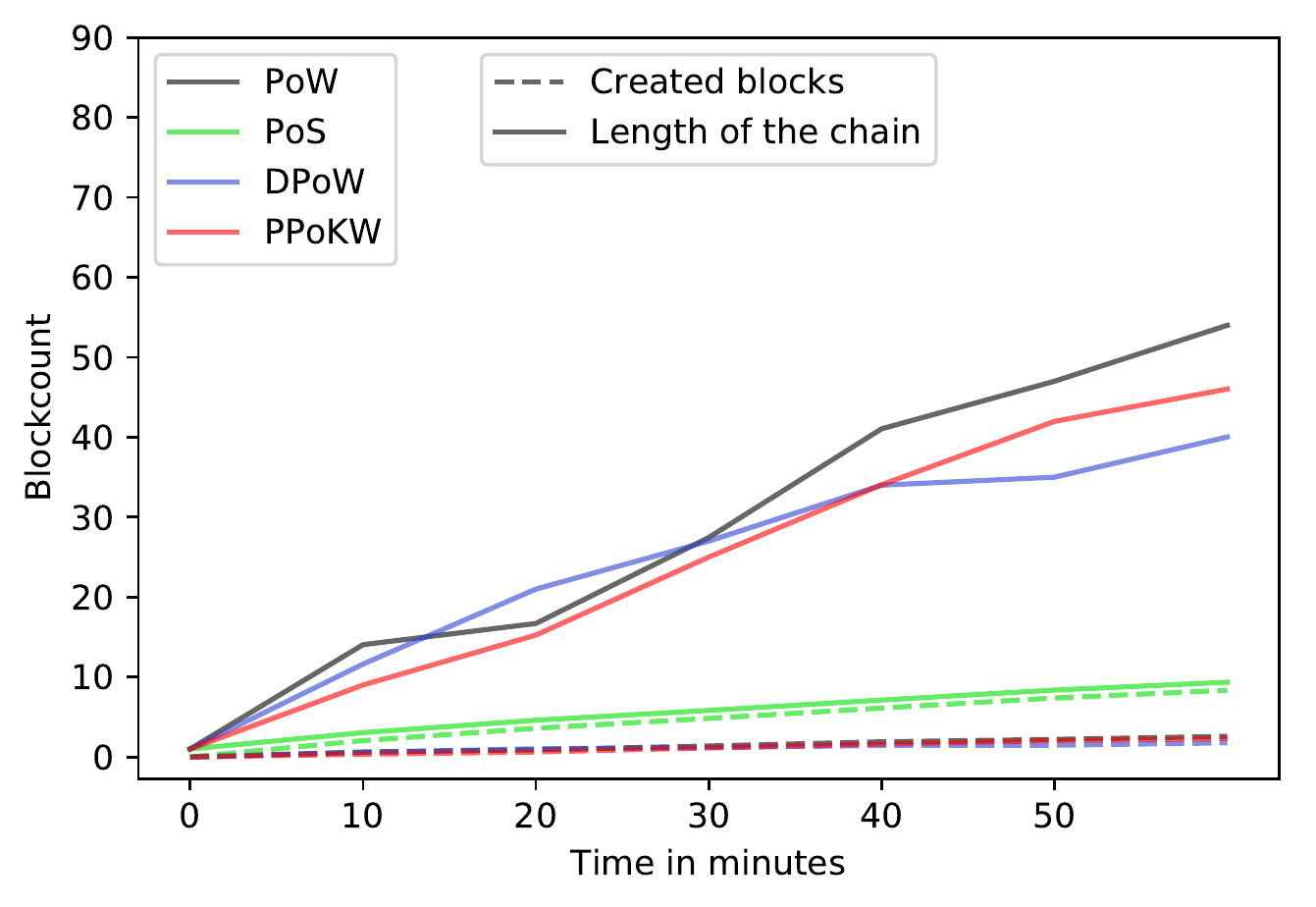}
        \caption{Average created/total blocks for 40 nodes over 1 hour}
        \label{fig:throughput_40}
    \end{minipage}        
\end{figure}
In Figures~\ref{fig:throughput_5} to ~\ref{fig:throughput_40} Proof-of-Stake stands out again. The time between consecutive mining Figures~\ref{fig:blocktime_5} to ~\ref{fig:blocktime_40} processes of one party could become quite high. Therefore broadcast of the transactions is important; otherwise, they would be available with a huge delay.

Next Figures~\ref{fig:memory_5} to ~\ref{fig:memory_40} depict the memory consumption of the methods.
\begin{figure}[h]
	\centering
    \begin{minipage}[t]{0.41\linewidth}
        \centering
        \includegraphics[width=\linewidth]{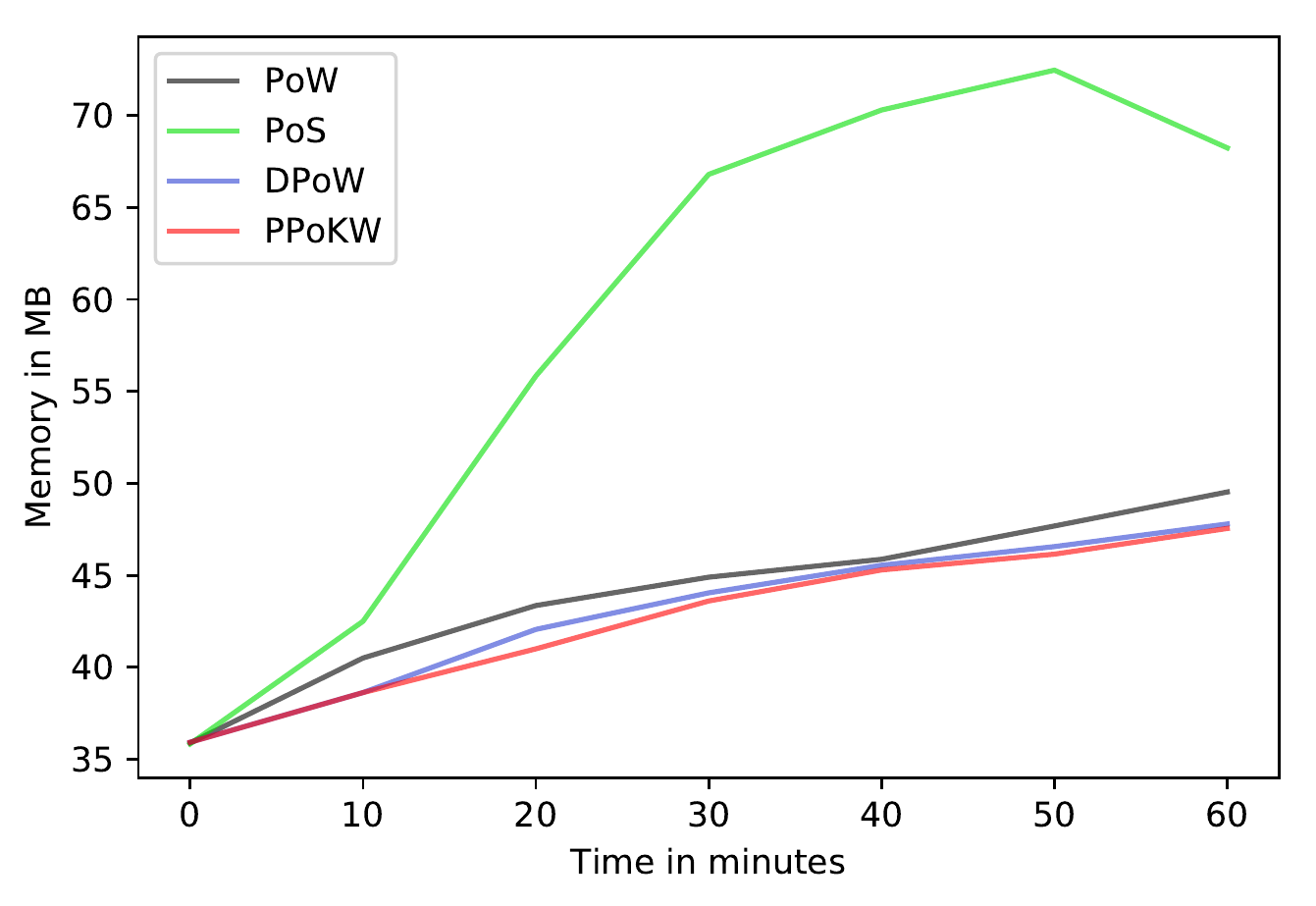}
        \caption{Average memory usage for 5 nodes over 1 hour}
        \label{fig:memory_5}
    \end{minipage}
    \hfill
    \begin{minipage}[t]{0.41\linewidth}
        \centering
        \includegraphics[width=\linewidth]{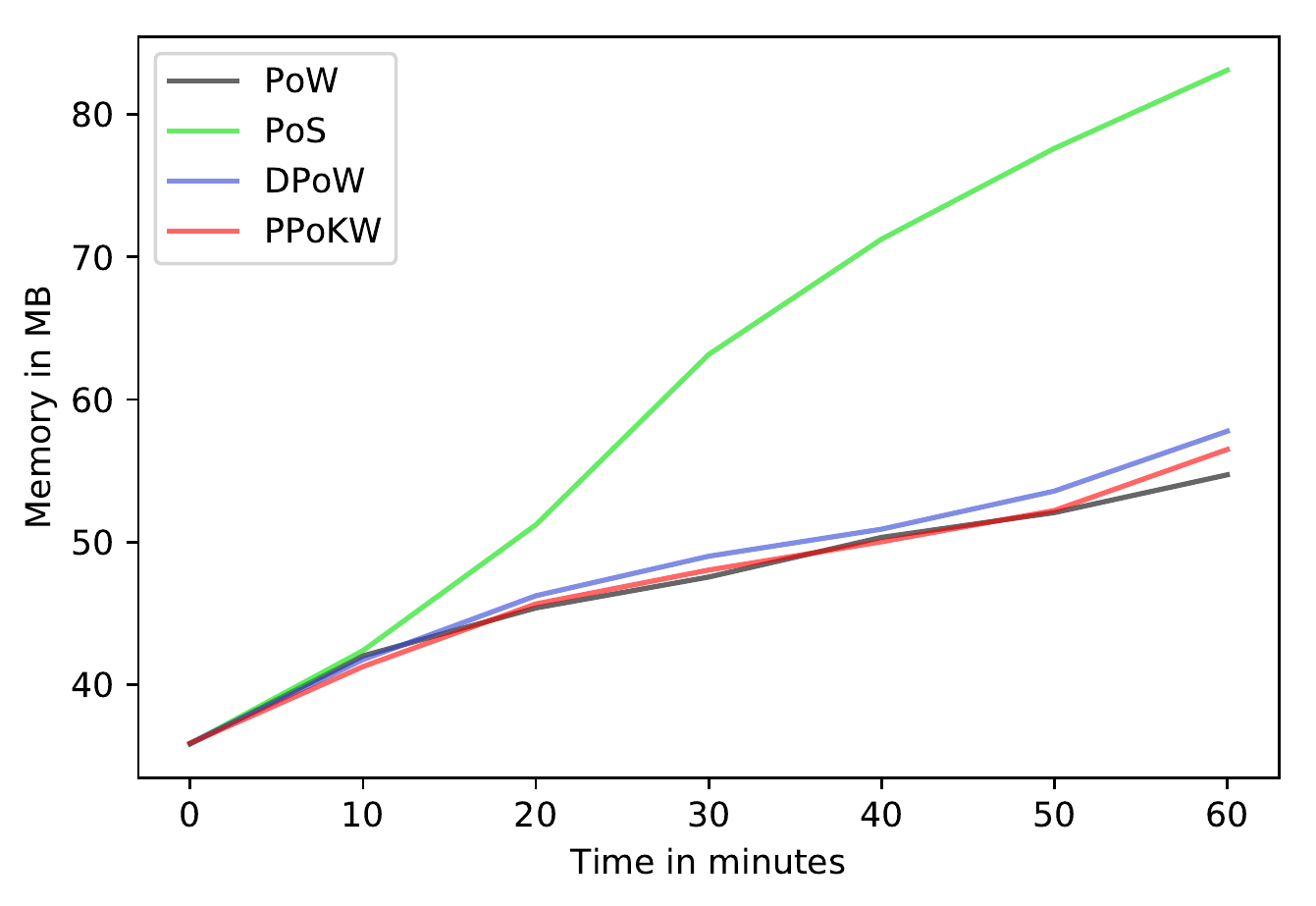}
        \caption{Average memory usage for 10 nodes over 1 hour}
        \label{fig:memory_10}
    \end{minipage}
	\centering
    \begin{minipage}[t]{0.41\linewidth}
        \centering
        \includegraphics[width=\linewidth]{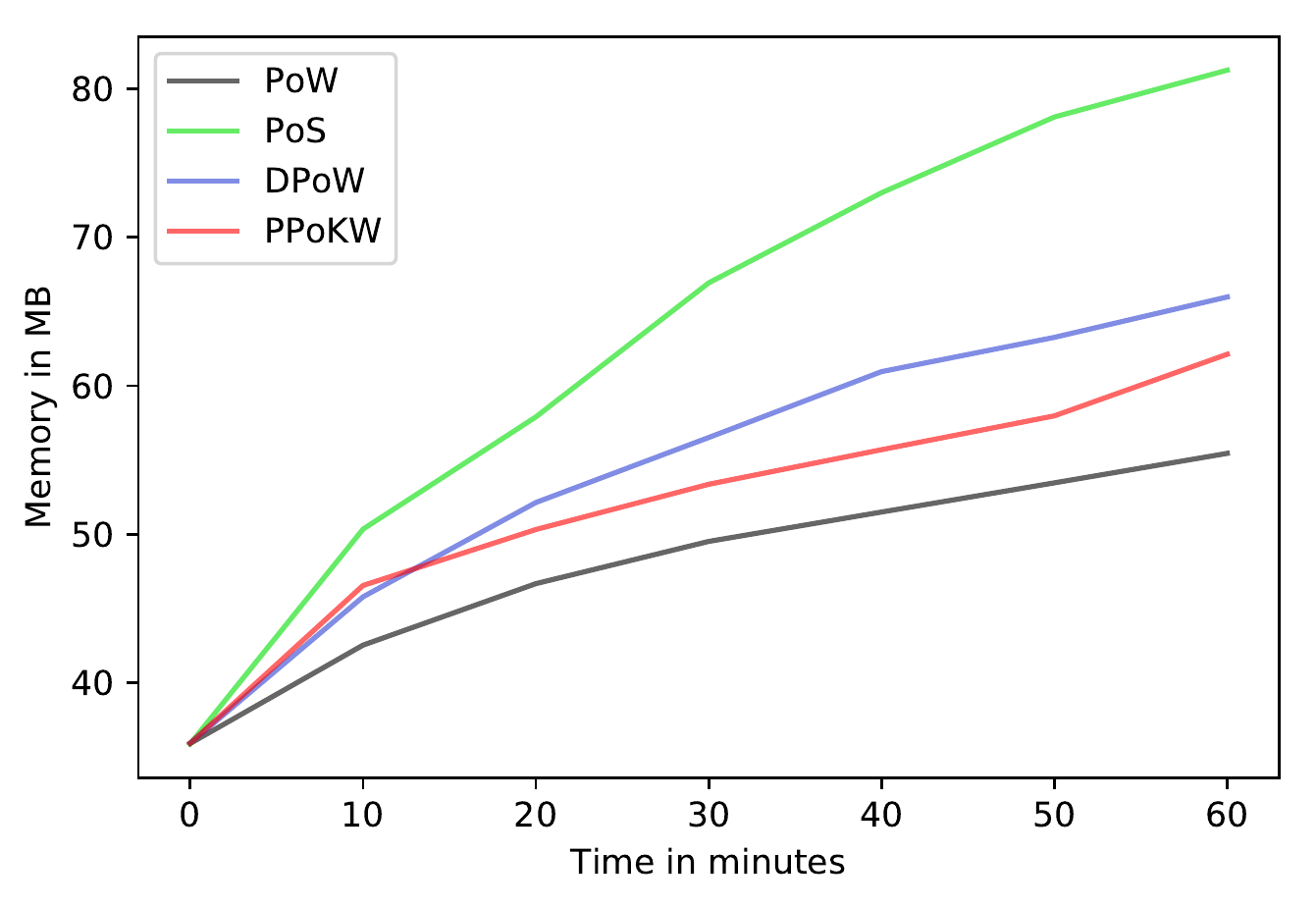}
        \caption{Average memory usage for 20 nodes over 1 hour}
        \label{fig:memory_20}
    \end{minipage}
    \hfill
    \begin{minipage}[t]{0.41\linewidth}
        \centering
        \includegraphics[width=\linewidth]{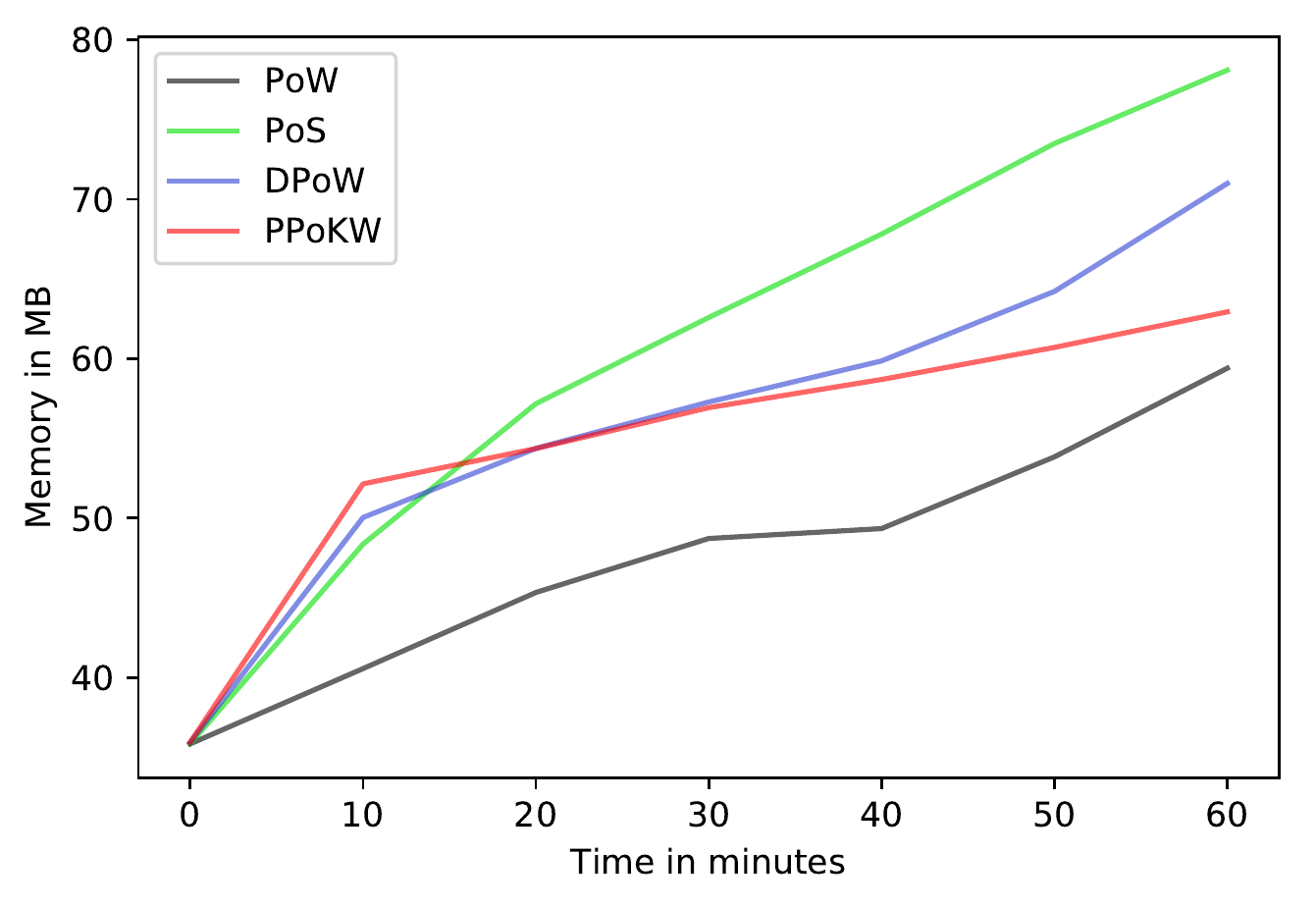}
        \caption{Average memory usage for 40 nodes over 1 hour}
        \label{fig:memory_40}
    \end{minipage}                 
\end{figure}
All four methods require a similar amount of memory. Proof-of-Work a little bit less, whereas the two methods with access control need a bit more memory.

\begin{figure}[h]
	\centering
    \begin{minipage}[t]{0.41\linewidth}
        \centering
        \includegraphics[width=\linewidth]{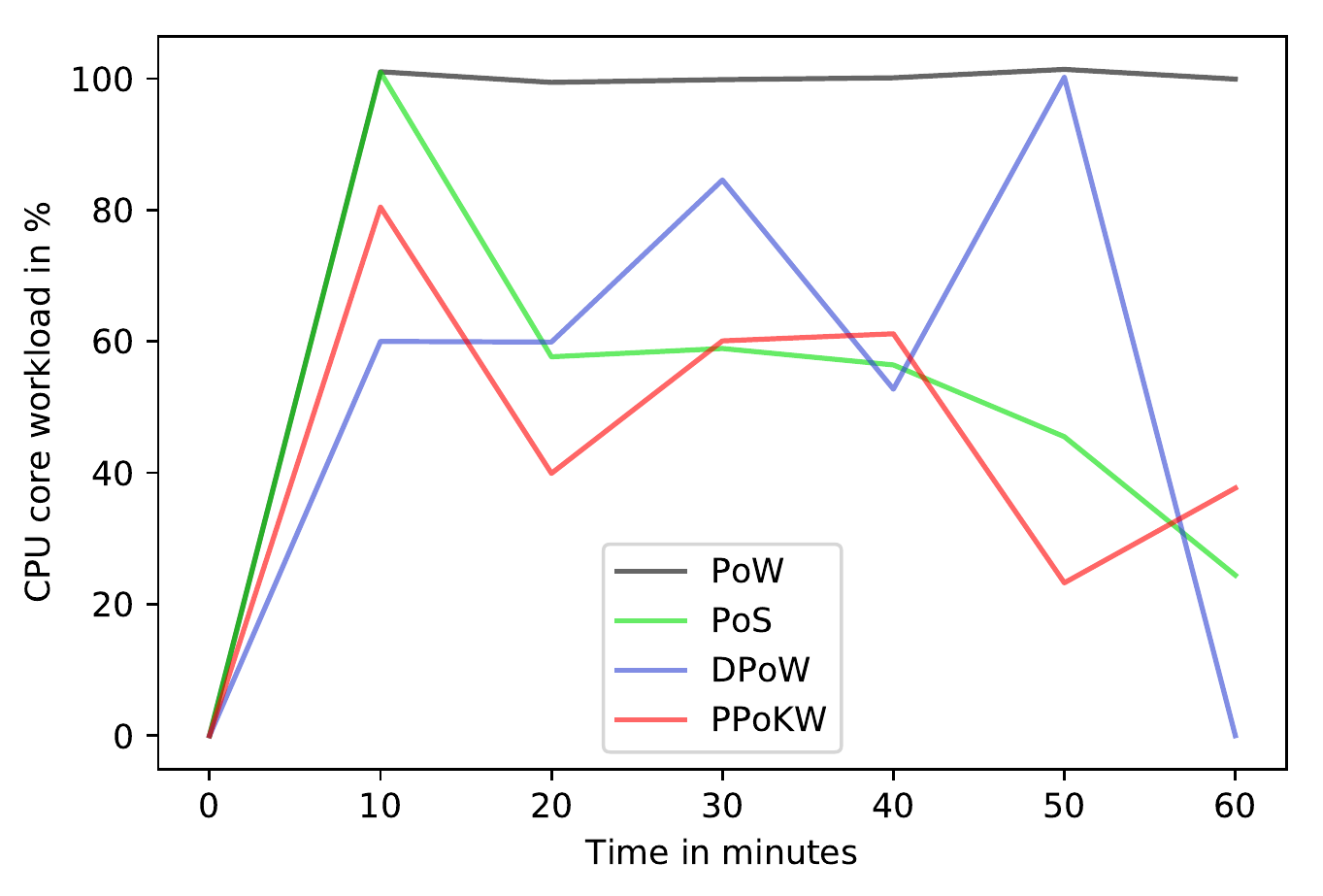}
        \caption{Average CPU-Workload for 5 nodes over 1 hour}
        \label{fig:cpu_5}
    \end{minipage}
    \hfill
    \begin{minipage}[t]{0.41\linewidth}
        \centering
        \includegraphics[width=\linewidth]{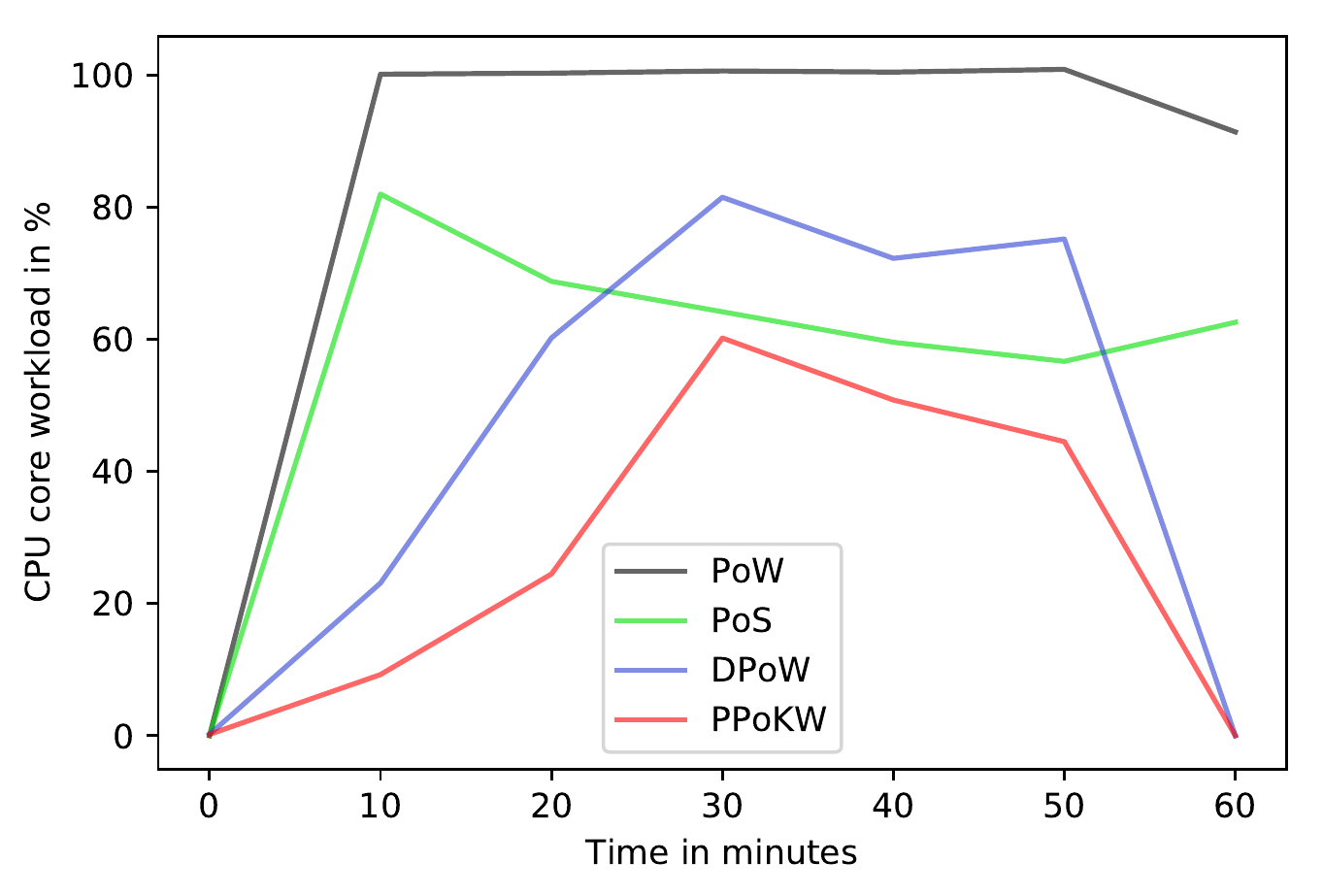}
        \caption{Average CPU-Workload for 10 nodes over 1 hour}
        \label{fig:cpu_10}
    \end{minipage}
	\centering
    \begin{minipage}[t]{0.41\linewidth}
        \centering
        \includegraphics[width=\linewidth]{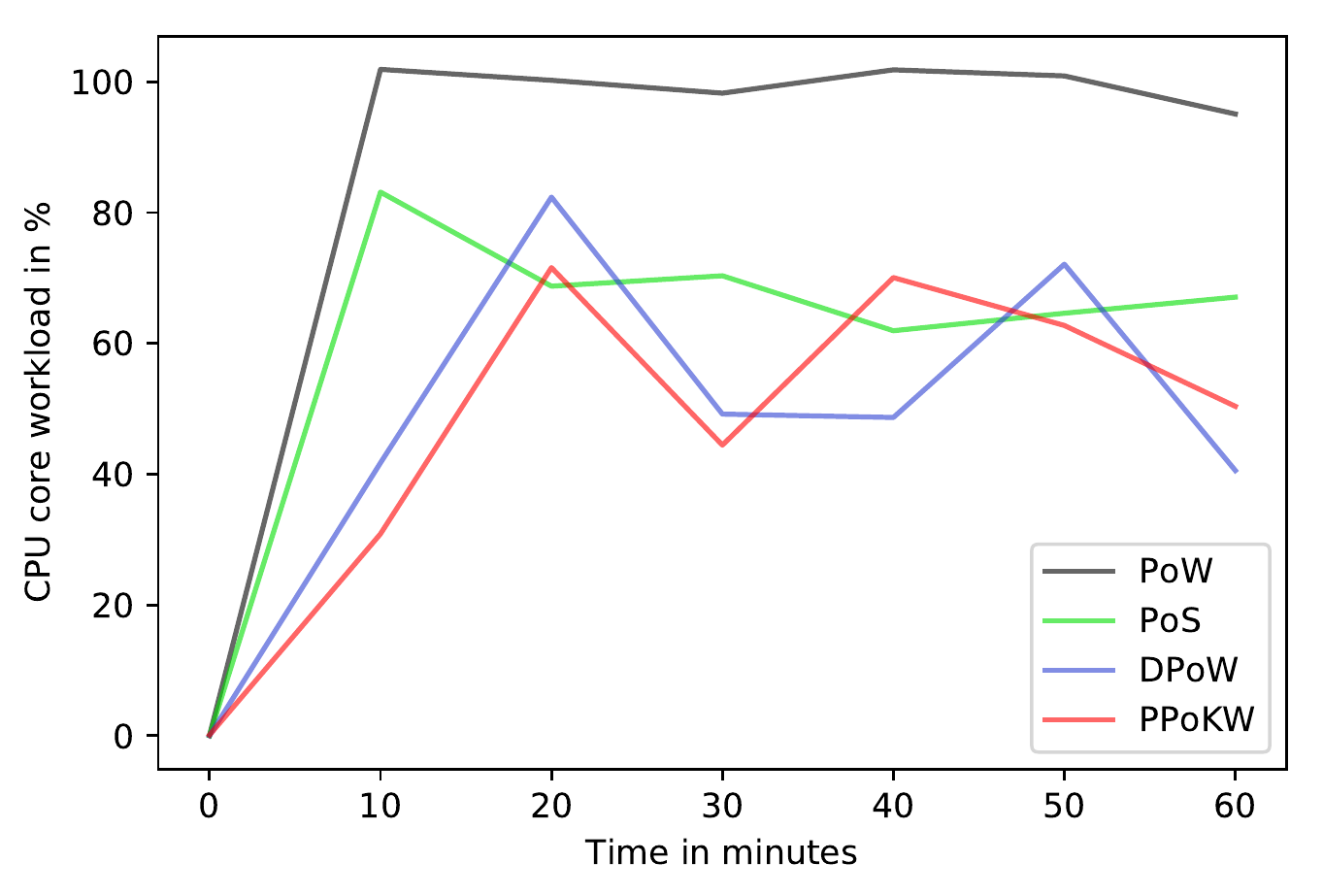}
        \caption{Average CPU-Workload for 20 nodes over 1 hour}
        \label{fig:cpu_20}
    \end{minipage}
    \hfill
    \begin{minipage}[t]{0.41\linewidth}
        \centering
        \includegraphics[width=\linewidth]{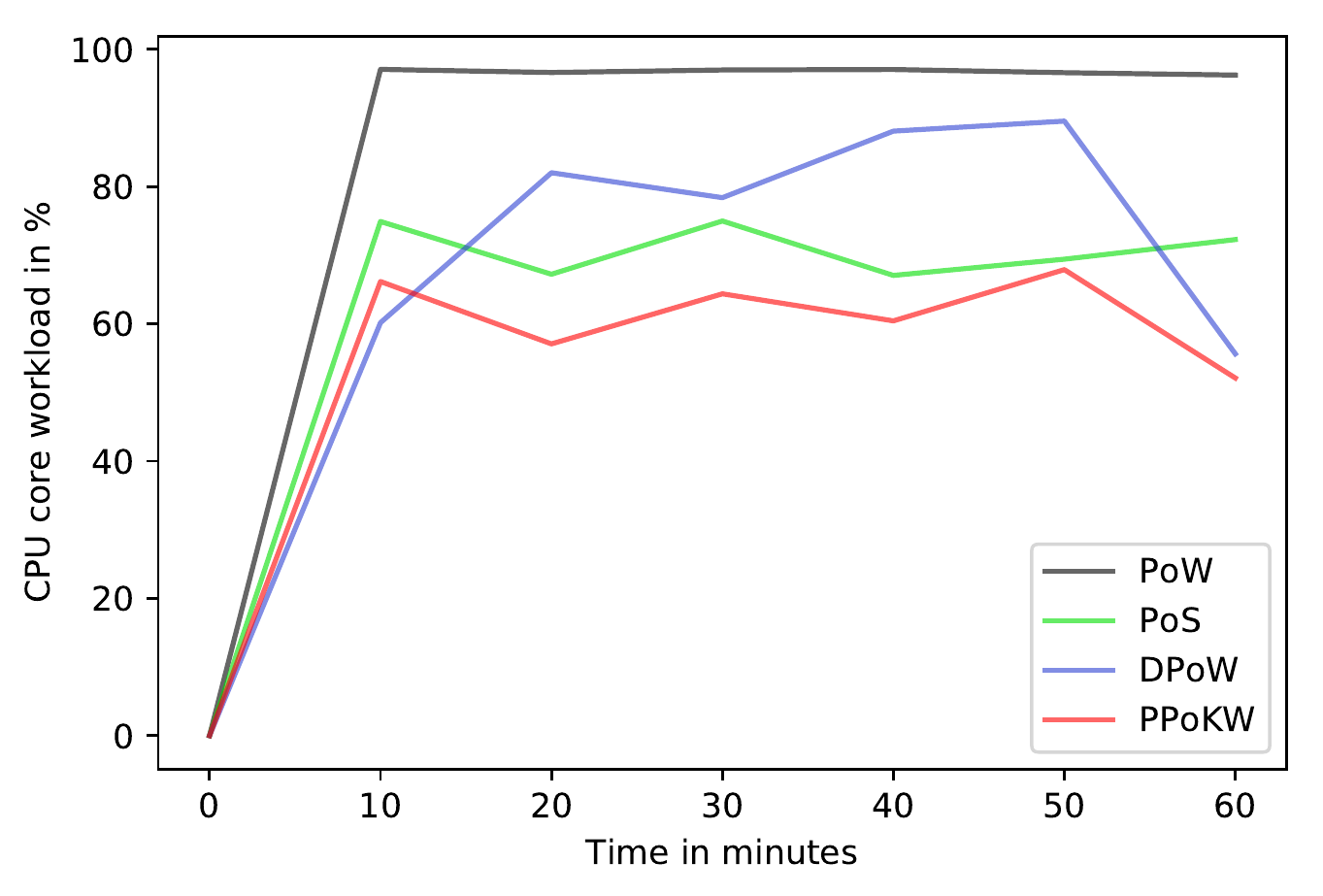}
        \caption{Average CPU-Workload for 40 nodes over 1 hour}
        \label{fig:cpu_40}
    \end{minipage}                 
\end{figure}

The CPU usage is depicted in Figures~\ref{fig:cpu_5} to ~\ref{fig:cpu_40}. Proof-of-Work requires a high percentage of the calculation power continuously. The two methods Practical Proof-of-Kernel-Work and Distributed Kernel-Work distribute the computation load more amongst the partners, and thus each CPU is used less. 

\section{Conclusion and Future Works}
The paper-at-hand assessed the suitability of blockchains to decentralized data processing scenarios. After we discussed promising consensus methods, we performed an event monitoring task. Thus we used a fully decentralized geometric monitoring. The analysis reveals that in cases of low data rates, where latencies by mining do not cause harm the methods could be integrated. A major drawback of blockchain is the requirement for broadcasts in the network. Besides communication costs, it also causes a blockchain on restricted memory parties to have a limit of participants given by the address list of the parties.
CPU usage does not cause a problem anymore as Distributed Proof-of-Work and Practical Proof-of-Kernel-Work overcome shortcomings of Proof-of-Work. 

The current analysis reveals that Proof-of-Work and Proof-of-Stake are not well suited for resource-constrained devices. In future work, the hardness and the verifiable random functions can be studied more. Also, investment strategies of coinage in combination with the restrictive consensus methods are promising.  

\FloatBarrier
\bibliography{paper}
\bibliographystyle{splncs04}
\end{document}